\newif\ifanon

\ifanon
  \documentclass[sigconf,anonymous,review]{acmart}
\else
  \documentclass[sigconf,nonacm]{acmart}
\fi
\renewcommand\footnotetextcopyrightpermission[1]{}
\AtBeginDocument{%
  }

\usepackage{pgfplots}
\pgfplotsset{compat=1.18}
\usetikzlibrary{calc}
\usepackage{comment}
\usepackage{xspace}
\usepackage{threeparttable}
\usepackage{booktabs}
\usepackage{tikz}
\usetikzlibrary{patterns}
\usepackage{cleveref}
\usetikzlibrary{positioning,arrows.meta,fit,backgrounds,calc}
\usetikzlibrary{decorations.markings}
\newcommand{\schemename}{\textsc{Sable}\xspace}

\newcommand{\cmark}{\ding{51}}
\newcommand{\xmark}{\ding{55}}
\newcommand{\pmark}{\ding{115}}
\usepackage{hyperref}

\begin{document}

\title{SABLE: Minimalist Instruction-Level Authenticated Encryption for Constrained Confidential Computing}

\ifanon
  \author{Anonymous Author(s)}
  \affiliation{%
    \institution{Submission under double-blind review}
    \country{}}
  \renewcommand{\shortauthors}{Anonymous et al.}
\else
  \author{Hamid Noori and Carlton Shepherd}
  \orcid{}                                  
  \affiliation{%
    \department{Department of Computer Science}
    \institution{Durham University}
    \city{Durham}
    \country{United Kingdom}}
  \email{\{hamid.noori, carlton.g.shepherd\}@durham.ac.uk}
  \renewcommand{\shortauthors}{Noori and Shepherd}
\fi

\begin{abstract}
Conventional processor designs expose code and data as plaintext throughout execution, rendering them inherently vulnerable to attacks that recover intellectual property or modify security/safety checks. Instruction-level encryption (ILE) enables CPU-level decryption, execution, and optionally authentication of individual encrypted program instructions at runtime. However, existing proposals depend on specific micro-architectures, detect corrupted instructions after they have executed, rely on non-standard ciphers, or require complex analyses of program state. In this work, we introduce and present a design exploration of a RISC-V processor architecture (\schemename{}) that enables minimally invasive instruction-level authenticated encryption of programs. \schemename{} is agnostic to the underlying micro-architecture, remaining compatible with the standard RISC-V toolchain with minor changes to post-process compiled ELF binaries. We integrate a decrypt-and-verify stage at two points---the instruction-memory wrapper and the CPU frontend---and explore seven ILE micro-architectures from a single-cycle (combinational) design to six multi-cycle (sequential) variants. We implement and evaluate the designs using ASCON-128a on a Xilinx Artix-7 FPGA with the open-source NEORV32 system on chip. Relative to baseline performance, the configurations span LUT, performance, power and energy per instruction overheads of $1.6$--$9.3\times$, $4.1$--$10\times$, $1.5$--$8\times$ and $10.4$--$80\times$, respectively, using the Dhrystone benchmarking suite. Finally, we discuss design trade-offs, highlighting area-, performance-, and energy-aware design points.
\end{abstract}



\maketitle

\section{Introduction}

Embedded processors form the core of cyber-physical systems across manufacturing~\cite{monjur2023hardware}, vehicle control units~\cite{milburn2018there}, medical devices~\cite{yaqoob2019security}, and industrial control systems~\cite{jiang2023monitoring}, where firmware performs safety-critical functions and embodies valuable intellectual property. Adversaries may target code memory to extract confidential firmware or to rewrite it so unauthorised code executes as if legitimate. This has been demonstrated through JTAG and debug-interface abuse, bus probing, and fault injection~\cite{shepherd2021physical,shwartz2017opening,vasile2018breaking}, and through bypasses of secure boot and read-out protection on Renesas~\cite{vandenherrewegen2021boots}, Volkswagen and BMW units~\cite{pozzobon2023fuzzy}, and the NVIDIA Tegra X2 used in the Tesla Autopilot and Mercedes-Benz infotainment systems~\cite{bittner2021voltage}. Firmware extraction, in turn, has enabled product cloning and counterfeiting in practice: some real-world examples include the Hondata s300 ECU~\cite{tehranipoor2017invasion} and Allen-Bradley CompactLogix programmable logic controllers (PLCs)~\cite{ulhaq2023survey,zhang2023aye}.

In response, \emph{instruction-level encryption} (ILE) conceals the instruction stream under a CPU-bound key~\cite{sinha2017reviving,gallagher2019morpheus,kleber2015eprint,papadogiannakis2013asist,hiscock2019lightweight}. Programs are provisioned in encrypted form and incrementally decrypted inside the processor, e.g., word by word, so attacks that successfully target program memory retrieve only encrypted instructions. Realisations differ in instruction set (ISA) and micro-architecture, cipher (e.g., PRINCE~\cite{lee2019ensuring}, XOR~\cite{papadogiannakis2013asist,lee2016binding}, Trivium~\cite{hiscock2019lightweight}), key management (e.g., PUF~\cite{kleber2015eprint}), and where the decryption boundary is drawn, whether at the fetch~\cite{hiscock2019lightweight} or decode~\cite{lee2019ensuring} stage, cache level~\cite{gallagher2019morpheus}, or MMU page~\cite{papadogiannakis2013asist}. For sensitive devices, the boundary should reside as deep inside the processor as can be afforded.


Addressing attacks that modify or inject new words is more difficult, wherein adversarial words may still decrypt to pseudo-random opcodes with undefined side-effects or ones which are valid.\footnote{Under XOR-based ILE schemes~\cite{papadogiannakis2013asist,lee2016binding}, ciphertexts are malleable; an attacker can modify targeted instruction bits without knowing the key.} One line of work tackles this by deriving the cipher state that decrypts each instruction from those executed beforehand~\cite{werner2018sponge,gousselot2025code}. These \emph{implicit} approaches still enable the execution of unauthorised garbage opcodes, however. They also require pre-computing inter-instruction states, necessitating deep modifications to toolchains and whole-program control-flow graph analyses, with restrictions to specific pipelines and ISAs~\cite{gousselot2025code,werner2018sponge}. Tackling this requires \emph{explicit} authentication, i.e., verifying code-associated authentication tags that raise an immediate exception upon mismatch.  Existing designs, however, amortise tags over blocks of instructions, restricting detection to block boundaries, and recover finer granularity only by using multiple layers of complexity~\cite{declercq2017sofia,savry2020confidaent}. More fundamental points in this space---per-instruction, explicit, chaining-free authenticated decryption from a single primitive with loose coupling to toolchains and micro-architectures---remain underexplored. 

In this paper, we present \schemename{} (Single-layer Authenticated, Block-free Lightweight Execution), a RISC-V architecture that enables minimalist authenticated ILE of independent instructions.   We realise \schemename{} on the NEORV32 core at two integration points---the instruction-memory (IMEM) wrapper and the CPU frontend---in encryption-only and authenticated configurations, exploring the design space across seven implementations, from a single-cycle (combinational) design to six multi-cycle variants. To our knowledge, \schemename{} is the first micro-architecture-agnostic, authenticated ILE architecture that characterises the design space across area, performance and energy consumption considerations. We further extend the Spike RISC-V ISA simulator \cite{riscvisasim} to support authenticated encrypted execution, providing a functional reference model for verifying and prototyping designs at the ISA level.

\subsection{Threat Model and Assumptions}
\begin{figure}
\centering
\resizebox{0.98\columnwidth}{!}{%
\begin{tikzpicture}[
  font=\small,
  >={Stealth[round]},
  box/.style={draw, semithick, rounded corners=1.5pt, align=center,
              minimum height=8mm, inner sep=3pt, fill=white},
  mem/.style   ={box, fill=gray!12, minimum width=15mm},
  crypto/.style={box, fill=blue!12, minimum width=27mm, minimum height=11mm},
  gate/.style  ={box, fill=blue!6, minimum width=22mm},
  pblk/.style  ={box, fill=gray!8, minimum width=17mm},
  io/.style    ={box, fill=white, font=\footnotesize, minimum width=16mm},
  flt/.style   ={box, fill=red!15, font=\footnotesize, minimum width=20mm},
  bus/.style   ={draw, semithick, fill=black!12, minimum width=80mm,
                 minimum height=5mm, inner sep=0pt, font=\footnotesize},
  arr/.style   ={->, semithick},
  darr/.style  ={<->, semithick},
  sig/.style   ={font=\footnotesize, inner sep=1.5pt},
  bnd/.style   ={font=\footnotesize\itshape},
  reg/.style   ={draw, semithick, rounded corners, inner sep=3mm},
]
\node[mem] (imem) {Encrypted\\IMEM};
\node[mem, right=7mm of imem] (tag)  {TAGMEM};
\node[mem, right=5mm of tag]  (sram) {Data\\SRAM};
\node[mem, right=5mm of sram] (io)   {I/O};
\coordinate (mmid) at ($(imem)!0.5!(io)$);
\node[bus, below=10mm of mmid] (bus) {System bus};
\node[crypto, below=11mm of bus, xshift=-26mm] (dec) {ASCON-128a\\decrypt + verify};
\node[io,  right=8mm of dec]     (key)  {key +\\nonce};
\node[gate, below=6mm of dec]    (gate) {\texttt{tag\_ok}?};
\node[flt, right=14mm of gate]   (fault){illegal instr.\\+ fault};
\node[pblk, below=8mm of gate]  (fe)   {Frontend};
\node[pblk, right=8mm of fe]    (ipb)  {IPB};
\node[pblk, right=8mm of ipb]   (be)   {Backend};
\draw[arr] (imem.south) -- (imem.south|-bus.north) node[sig,pos=0.55]{};
\draw[arr] (tag.south|-bus.north) -- (tag.south);
\draw[arr] (bus.south -| dec) -- (dec.north)
           node[sig,left, pos=0.49,xshift=-1mm]{ciphertext};
\draw[arr] (key)  -- (dec);
\draw[arr] (dec)  -- node[sig,right,xshift=1mm]{plaintext} (gate);
\draw[arr] (gate) -- node[sig,above]{fail} (fault);
\draw[arr] (gate.south) -- node[sig,right,pos=0.4,xshift=1mm]{pass} (fe.north);
\draw[arr] (fe)  -- (ipb);
\draw[arr] (ipb) -- (be);
\draw[darr] (sram.south) -- (sram.south|-bus.north);
\draw[darr] (io.south)   -- (io.south|-bus.north);
\draw[darr] (bus.south -| be) -- node[sig,right,pos=0.85,xshift=1mm]{data / MMIO} (be.north);
\begin{scope}[on background layer]
  \node[reg, dashed, fill=red!4,
        fit=(imem)(tag)(sram)(io)(bus),
        label={[bnd]above:Untrusted memory and bus}] (ureg) {};
  \node[inner sep=3mm, fit=(key)(dec)(gate)(fault)(fe)(ipb)(be)] (tcont) {};
  \node[draw, semithick, rounded corners, fill=green!4, inner sep=0,
        fit=(ureg.west |- tcont.north)(ureg.east |- tcont.south),
        label={[bnd]below:Trusted core}] {};
  \node[draw, dotted, semithick, rounded corners, inner sep=2.2mm,
        fit=(fe)(ipb)(be),
        label={[bnd]above:NEORV32 pipeline}] {};
  \draw[dashed,arr] (tag.west) -| ($(dec.north)+(5mm,0)$) node[sig, xshift=3.3mm,pos=0.925, above]{tag*};
\end{scope}
\end{tikzpicture}%
}
\caption{\schemename{} block diagram (\texttt{FEAUDC}~design; \Cref{sec:frontend-integration}). \footnotesize{(*\texttt{TAGMEM} is directly connected to the ASCON module, not via the system bus.)}}
\label{fig:overview2}
\end{figure}
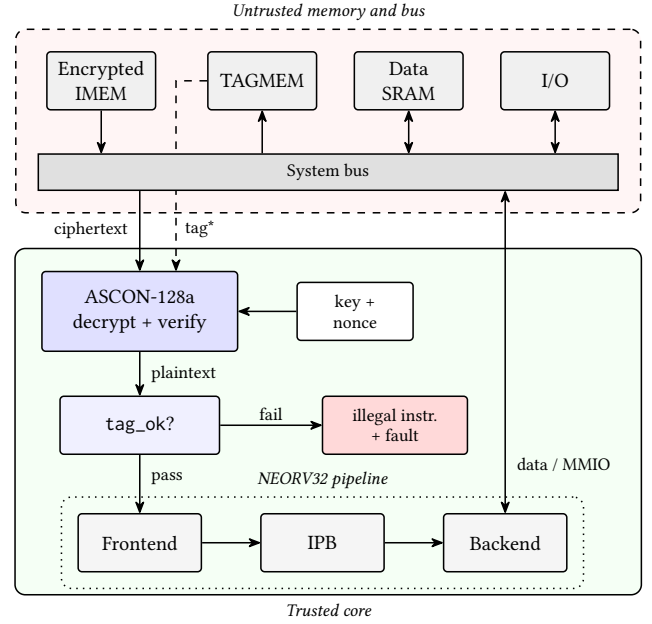

We consider a conventional system-on-chip with a single processor, optionally including instruction and data caches. The processor core is considered to be trusted, encompassing the internal state, e.g., program counter, register file, and arithmetic logic units, where the execution is performed. We consider attackers capable of observing and modifying instructions stored in untrusted memory, e.g., IMEM, whether partially or in full, alongside bus signals and memory interfaces; and inject new unauthorised instructions into the instruction stream in untrusted elements (e.g., at the system bus). Attack scenarios include extracting program instructions to counterfeit firmware and compromise intellectual property, performing unauthorised code execution or modifying firmware in order to disable security and safety controls.  We define the trust boundaries in Figure~\ref{fig:overview2}. Physical or micro-architectural attacks that target this region are thus considered out of scope. \schemename{} focusses on instruction-level protection; control-flow integrity (CFI) of higher-level program semantics is not covered explicitly by \schemename{}. Code reuse attacks are similarly not covered, e.g., return-to-libc~\cite{tran2011expressiveness} and ROP~\cite{roemer2012rop} and JOP~\cite{bletsch2011jump} gadgets in the (encrypted) binary. 

\subsection{Related Work}

Hardware ILE schemes divide into encryption-only and authenticated designs; Table~\ref{tab:related} positions \schemename{} against both groups by security and deployability considerations.

\emph{Encryption-only designs.}  Polyglot \cite{sinha2017reviving} combines AES and elliptic curve cryptography for page-level encryption, extending protection to OSs and hypervisors; however, its reliance on MMUs limits applicability to lightweight embedded systems. Morpheus \cite{gallagher2019morpheus} protects memory contents between cache levels using the QARMA cipher \cite{avanzi2017qarma} but is evaluated only via gem5 simulation \cite{binkert2011gem5}. ASIST \cite{papadogiannakis2013asist} integrates hardware support for ILE at instruction and block granularity using XOR and transposition, with a later extension introducing AES-based block decryption before the instruction cache \cite{christou2020architectural}. A scheme based on the Trivium stream cipher \cite{hiscock2019lightweight} decrypts instructions during fetching, ensuring plaintext exists only within the core. Programs are divided into basic blocks with independent initialisation vectors (IVs) for greater granularity, though integrity verification is absent. PolEn \cite{morel2023polymorphism} extends this with runtime polymorphism to mitigate side-channel analysis, while Morpheus II \cite{harris2021morpheusii} tackles control-flow hijacking protection at instruction granularity on RISC-V. ECP \cite{gao2025ecp} uses PRINCE-based encryption for program consistency but lacks hardware validation. Lee et al.\ chain an LFSR mask through previously stored instructions to bind firmware to a device against counterfeiting~\cite{lee2016binding}; their later SEP uses PRINCE-encrypted basic blocks to the block address and pairs \texttt{CALL}/\texttt{RETURN} through explicit check instructions~\cite{lee2019ensuring}, gaining coarse-grained integrity at the cost of a custom ISA and assembler.

\emph{Authenticated designs.} Explicit two-layer schemes pair encryption with a separate integrity mechanism at block granularity: SOFIA~\cite{declercq2017sofia} verifies CBC-MACs over blocks of four to six instructions, and CONFIDAENT~\cite{savry2020confidaent} applies ASCON with 64-bit tags to 16-byte blocks plus mask-based control-flow integrity (CFI) requiring 256 bits of metadata per block. Both schemes confine detection to block boundaries and require dedicated compiler support. Implicit single-layer schemes instead chain the cipher state through the instruction stream: SCFP~\cite{werner2018sponge} places a sponge-based decrypt stage between fetch and decode, and Gousselot et al.~\cite{gousselot2025code} extend this philosophy with an ASCON duplex in the CV32E40P fetch stage that binds the pipeline's control signals. Chaining buys CFI at low cost (SCFP reports $9.1\%$ runtime and $19.8\%$ code size~\cite{werner2018sponge}), but detection lacks determinism: no tag is generated, and a tampered word is caught only once the diverging keystream decodes into an illegal opcode, which may be after tens of instructions~\cite{gousselot2025code}. Chaining schemes necessitate deep coupling with program semantics: whole-program control-flow graph analysis, per-edge patching or ISA extensions, and pipeline-specific control-signal modelling~\cite{werner2018sponge,gousselot2025code}. Ahmad et al.~\cite{ahmad2025chaining} use explicit per-instruction HMACs to prevent replay and control-flow hijacking, but combines separate encryption and authentication layers, and it is not evaluated in hardware.

\begin{table*}
\centering
\caption{Positioning of \schemename{} against related
hardware-assisted instruction-protection schemes.}
\label{tab:related}
\footnotesize
\resizebox{\textwidth}{!}{%
\begin{tabular}{@{}r|llllccccl@{}}
\toprule
\textbf{Scheme} & \textbf{Primary threat} & \textbf{Primitive} & \textbf{Granularity} & \textbf{ISA} & \textbf{Auth.} & \textbf{\shortstack{Pre-exec.\\detection}} & \textbf{\shortstack{Single\\layer}} & \textbf{\shortstack{Analy.\\-free}} & \textbf{Validation} \\
\midrule
ASIST~\cite{papadogiannakis2013asist}   & Code injection                 & XOR / transposition           & Instr.\ / block (enc.) & SPARC V8              & \xmark & \xmark & \cmark & \cmark & FPGA        \\
Morpheus~\cite{gallagher2019morpheus}   & Control-flow hijacking   & QARMA-64 + displacement       & 64-bit word (enc.)     & RISC-V                & \xmark & \xmark & \xmark & \pmark & Sim.\ (gem5) \\
Morpheus II~\cite{harris2021morpheusii} & Remote code exec.       & SIMON (12-round)              & Instr.\ + code ptr.\ (enc.) & \textsuperscript{\ding{91}}RISC-V & \xmark & \xmark & \cmark & \xmark & FPGA        \\
Lee et al.~\cite{lee2016binding}           & Counterfeiting, IP theft       & Chained LFSR masking          & Instruction (enc.)     & PicoBlaze         & \xmark & \xmark & \cmark & \cmark & FPGA \\
SEP~\cite{lee2019ensuring}          & Tampering, counterfeiting      & PRINCE (addr.-keyed, chained) & Block (4 instr.)       & Custom           & \pmark & \pmark & \xmark & \xmark & FPGA \\
SOFIA~\cite{declercq2017sofia}          & Code tampering, reuse          & Enc.\ + CBC-MAC               & Block (4--6 instr.)    & SPARC V8              & \cmark & \cmark & \xmark & \xmark & FPGA        \\
CONFIDAENT~\cite{savry2020confidaent}   & Faults, code reuse             & ASCON + masking               & Block (16\,B)          & \textsuperscript{\ding{91}}RISC-V & \cmark & \cmark & \xmark & \xmark & FPGA        \\
SCFP~\cite{werner2018sponge}            & Faults, code reuse             & Sponge AE (chained)           & Instruction            & \textsuperscript{\ding{91}}RISC-V & \pmark & \pmark & \cmark & \xmark & ASIC        \\
Gousselot et al.~\cite{gousselot2025code}  & Faults, control signals        & ASCON duplex (chained)        & Instruction            & RISC-V                & \pmark & \pmark & \cmark & \xmark & FPGA        \\
Ahmad et al.~\cite{ahmad2025chaining}      & Code tampering                 & Enc.\ + HMAC (chained)        & Instruction            & RISC-V (sim.)         & \cmark & \cmark & \xmark & \xmark & Python  \\
\midrule
\textbf{\schemename{}}      & Injection, tampering, IP theft & ASCON (Design space exploration)              & Instruction            & RISC-V                & \cmark & \cmark & \cmark & \cmark & FPGA \\
\bottomrule
\end{tabular}}
\par\smallskip
\begin{minipage}{\textwidth}
\footnotesize\raggedright
\textit{Selected column details.} 
\textbf{Granularity}: the smallest unit of operation for encryption and authentication.
\textbf{Auth.}: cryptographic authentication is provided.
\textbf{Pre-exec.~detection}: unauthorised instructions are detected prior to execution.
\textbf{Single layer}: a single cryptographic mechanism provides all
claimed protection.
\textbf{Analy.-free}: no program semantic analysis, binary patching, or ISA modification is required. \textit{Symbols.} \cmark~= provided; for detection, deterministic and before the instruction executes. \pmark~= partial or implicit; detection yields pseudo-random opcodes
that may execute. \xmark~= absent.
\ding{91}~= Requires non-standard instructions added to the base ISA and corresponding compiler support.
\end{minipage}
\end{table*}

\emph{Positioning.} \schemename{} uses a single authenticated encryption with associated data (AEAD) primitive with instruction independence, so a failed verification raises a deterministic hardware fault in the fetch cycle, rather than after several pseudo-random instructions may have retired. \schemename{} does not require CFG analysis, binary patching, nor compiler modification besides a modified \texttt{objcopy}, and ports across micro-architectures. We do not claim to exceed other scheme' performance: per-instruction tags cost \schemename{} $4.1$--$10\times$ runtime and $5\times$ IMEM where SCFP costs ${\approx}9\%$ and ${\approx}20\%$~\cite{werner2018sponge}. That reflects a 64-bit PRINCE state; instantiated with ASCON's 320-bit state, the same chained construction costs roughly $10\times$ IMEM in patches and a $30$--$72\%$ clock-frequency reduction~\cite{gousselot2025code}, placing \schemename within the established cost of ASCON-class protection. We consider this appropriate where deterministic pre-execution detection and toolchain independence are preferred. Nor do we claim CFI, which \schemename{} deliberately forgoes.  The hardware cost space of this approach remains largely uncharted, which we address, and no prior scheme reports power or energy, or integration points.

\section{Proposed Designs}
We now explore the design space of integrating authenticated ILE into an embedded processor. First, single-cycle implementations of ASCON-128a decryption are presented, with and without authentication, and with their integration into IMEM and CPU frontend of the reference (NEORV32) architecture. Second, multi-cycle implementations are discussed, and their impact on performance, timing, energy and system integration.

\subsection{Single-Cycle Decryption Architectures}\label{subsec:single}

\schemename{} employs a lightweight cryptography module introduced, firstly, at the instruction fetch stage. ASCON-128a operates on a 320-bit internal state of five 64-bit words in a sponge construction; its core permutation combines round-constant addition and a 5-bit substitution layer; the algorithm takes a 128-bit key and 128-bit public nonce and produces a 128-bit authentication tag. We develop a single-cycle VHDL implementation of ASCON-128a decryption specialised for a fixed 32-bit ciphertext (one RISC-V instruction) and zero-length associated data. Under these constraints, this enables a fully unrolled, straight-line combinational datapath organised as (i) an initialisation stage loading the IV, key, and nonce into the state; (ii) a fully unrolled 12-round \texttt{ASCON-p[12]} permutation; and (iii) a decryption stage producing the plaintext instruction. To synthesise the fully combinational permutation, it is modularised into a dedicated VHDL entity with explicit inter-round signals annotated \texttt{DONT\_TOUCH}, preventing aggressive optimisation. A tagless (confidentiality-only) variant is also explored.

For authentication, a 128-bit tag is produced and compared with the expected one, yielding a binary signal (\texttt{tag\_ok}). If it fails, the decrypted output is replaced with an illegal instruction and a hardware exception is asserted, preventing the execution of unauthorised instructions. The full 128-bit ASCON-128a tag is retained per instruction word; this memory cost is a deliberate trade-off in favour of an unreduced integrity margin rather than a consequence of the tag size. This requires $5\times$ the instruction memory capacity compared to the baseline processor and the design employing tagless decryption. 
For this baseline, we use NEORV32~\cite{nolting2022neorv32}, a compact 32-bit RISC-V soft-core with a multi-cycle, finite-state machine (FSM)-based frontend (instruction fetch and buffering) and backend (decode and execution), decoupled by an Instruction Prefetch Buffer (IPB), with coupled instruction and data memories on a memory-mapped bus. We integrate the decryptor at two points: the IMEM and the CPU frontend.

\subsubsection{IMEM-Level Integration}

In our baseline design (\texttt{IMDEC}), a single-cycle tagless decryptor is inserted into the IMEM read path (\texttt{neorv32\_imem.vhd}), transparently decrypting all instruction and read-only-data accesses within a single clock cycle, before they reach the processor bus and with no protocol changes. The key and a base nonce are stored in dedicated write-only registers; the per-instruction nonce is formed by adding the 32-bit fetch address into the lowest 32 bits of the 128-bit base nonce, binding each ciphertext to its memory location so that a valid (ciphertext, tag) pair cannot be relocated or spliced to a different address (Eq.~\ref{eq:nonce-deriv}).

\begin{equation}
    N(a) \;=\; N_{\mathrm{base}}[127{:}32] \;\Vert\; \bigl(\,(N_{\mathrm{base}}[31{:}0] + (a \gg 2)) \bmod 2^{32}\bigr)
    \label{eq:nonce-deriv}
\end{equation}

This approach offers minimal architectural intrusion at the cost of critical-path delay. The timing path spans the memory array, the full combinational ASCON datapath, and the bus output, limiting the maximum operating frequency. We also develop an authenticated variant (\texttt{IMAUDC}) that adds a dedicated Tag Memory (TAGMEM)---a synthesis-time-initialised VHDL package storing one 128-bit tag/output per instruction word, accessed synchronously with IMEM under the same address and directly connected to the cryptographic module. TAGMEM's contents are not confidential: because its entries are ASCON-128a tags, an adversary cannot produce a matching tag for a chosen instruction without the key. The tags may reside in the same untrusted memory as the encrypted code as an optional design consideration. The bus is driven with the plaintext instruction only when \texttt{tag\_ok} holds; otherwise an illegal instruction is injected and a bus error raised, preventing execution of tampered code without modifying the processor core.

\subsubsection{Frontend-Level Integration}
\label{sec:frontend-integration}

We develop an encryption-only design variant (\texttt{FEDEC}) that places the decryptor between the instruction fetch and IPB stages. The frontend operates as an FSM with three states: \texttt{RESTART}, \texttt{REQUEST}, and \texttt{PENDING}, managing fetching and control-flow changes. In a second authenticated design variant (\texttt{FEAUDC}), we add a frontend TAGMEM indexed by the fetch address, whose 128-bit output feeds an ASCON module directly. Only instructions with a valid tag are written into the IPB; otherwise an illegal instruction is injected and an instruction-fetch error raised. Frontend integration limits the exposure of decrypted instructions and provides fine-grained control over instruction validation while maintaining compatibility with the existing NEORV32 pipeline.

\subsection{Multi-Cycle Architectures}

The single-cycle authenticated design instantiates two ASCON-$p[12]$ permutations---one for decryption, one for tag generation and verification---resulting in a long critical path and high area cost. Our multi-cycle family addresses this by folding: partitioning ASCON-$p[12]$ into $r$ rounds per cycle and reusing the logic across cycles shortens the critical path and shares resources, at the cost of a higher cycle count per instruction (CPI). Six designs, \texttt{MCP12} through \texttt{MCP1}, implement 12 down to 1 round(s) per cycle; \Cref{tab:mcp} summarises the resulting permutation latency, FSM complexity, and added instruction latency. All designs share a common interface (\texttt{rstn}, \texttt{start}, \texttt{clk}, \texttt{done}) with input and output registers for ciphertext, plaintext, and \texttt{tag\_ok}; an FSM sequences the time-multiplexed permutation passes, split evenly between decryption and authentication.  The \texttt{rstn} signal initialises the internal FSM to an \texttt{IDLE} state, while assertion of \texttt{start} triggers the decryption and authentication process. Upon tag verification, the \texttt{done} signal is asserted for one clock cycle.  The multi-cycle decryptors are integrated into the NEORV32 CPU frontend by modifying its three-state fetch FSM (\texttt{RESTART}, \texttt{REQUEST}, \texttt{PENDING}): \texttt{RESTART} asserts the decryptor reset; \texttt{PENDING} loads the acknowledged ciphertext into the decryptor's input register and asserts \texttt{start}; and a new wait state holds until \texttt{done} is asserted, upon which the plaintext is written to the IPB if \texttt{tag\_ok} holds, otherwise an illegal instruction machine code is written to the IPB and an exception is triggered. The frontend updates the program counter and resumes normal control flow. This integration enables a trade-off between performance and hardware cost, allowing designers to select an appropriate architecture based on application constraints.

\begin{table}[t]
\caption{Multi-cycle \schemename{} designs.}
\label{tab:mcp}
\begin{center}
\vspace{-3mm}
\footnotesize
\begin{tabular}{lcccccc}
\toprule
 & \texttt{MCP12} & \texttt{MCP6} & \texttt{MCP4} & \texttt{MCP3} & \texttt{MCP2} & \texttt{MCP1} \\
\midrule
ASCON rounds / cycle          & 12 & 6 & 4 & 3 & 2 & 1 \\
Cycles / ASCON perm.    & 1  & 2 & 3 & 4 & 6 & 12 \\
Time-mux.\ cycles       & 2  & 4 & 6 & 8 & 12 & 24 \\
FSM states              & 5  & 7 & 9 & 11 & 15 & 27 \\
Added latency per inst. (cycles)  & 7  & 9 & 11 & 13 & 17 & 29 \\
\bottomrule
\end{tabular}
\end{center}
\end{table}

\section{Toolchain and Verification}\label{sec:toolchain}
 

 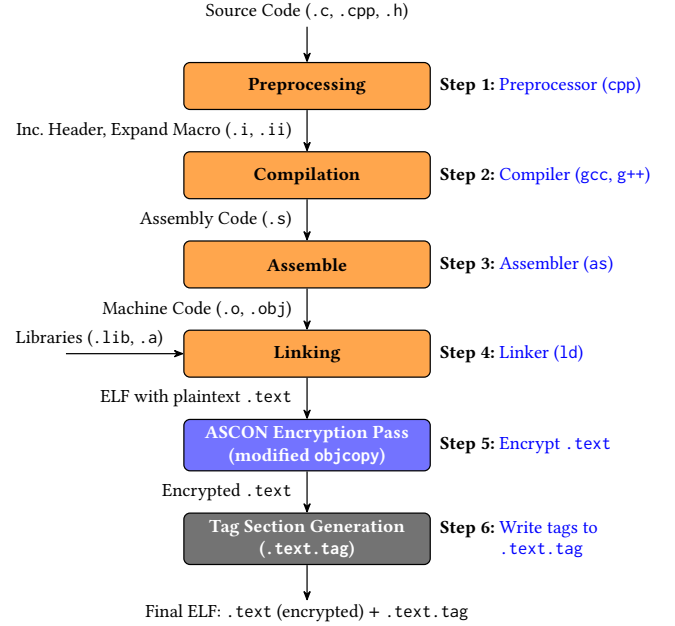
\begin{figure}[t]
\centering
\resizebox{\columnwidth}{!}{%
\begin{tikzpicture}[
    font=\normalsize,
  >={Stealth[round]},
  stage/.style={draw, semithick, rounded corners=3pt, align=center,
                fill=orange!70, text=black, font=\normalsize\bfseries,
                minimum width=42mm, minimum height=8mm, inner sep=3pt},
  secstage/.style={draw, semithick, rounded corners=3pt, align=center,
                fill=blue!55, text=white, font=\normalsize\bfseries,
                minimum width=42mm, minimum height=8mm, inner sep=3pt},
  thirdstage/.style={draw, semithick, rounded corners=3pt, align=center,
                fill=black!55, text=white, font=\normalsize\bfseries,
                minimum width=42mm, minimum height=8mm, inner sep=3pt},
  lbl/.style={font=\normalsize, align=right, inner sep=1.5pt},
  stp/.style={font=\normalsize, align=left, inner sep=1.5pt},
  arr/.style={->, semithick, >={Stealth[round]}},
]
 
\node[stage] (pre)  {Preprocessing};
\node[stage,    below=7mm of pre]  (comp) {Compilation};
\node[stage,    below=7mm of comp] (asm)  {Assemble};
\node[stage,    below=7mm of asm]  (link) {Linking};
\node[secstage, below=7mm of link] (enc)  {ASCON Encryption Pass\\(modified \texttt{objcopy})};
\node[thirdstage, below=7mm of enc]  (tag)  {Tag Section Generation\\(\texttt{.text.tag})};
 
\draw[arr] ($(pre.north)+(0,6mm)$) -- (pre.north)
    node[lbl, pos=0, above=0.5mm, align=left]
        {Source Code (\texttt{.c}, \texttt{.cpp}, \texttt{.h})};
\draw[arr] (pre.south)  -- (comp.north)
    node[lbl, midway, left, xshift=-2mm]
        {Inc.~Header, Expand Macro (\texttt{.i}, \texttt{.ii})};
\draw[arr] (comp.south) -- (asm.north)
    node[lbl, midway, left, xshift=-2mm]
        {Assembly Code (\texttt{.s})};
\draw[arr] (asm.south)  -- (link.north)
    node[lbl, midway, left, xshift=-2mm]
        {Machine Code (\texttt{.o}, \texttt{.obj})};
\draw[arr] (link.south) -- (enc.north)
    node[lbl, midway, left, xshift=-2mm]
        {ELF with plaintext \texttt{.text}};
\draw[arr] (enc.south)  -- (tag.north)
    node[lbl, midway, left, xshift=-2mm]
        {Encrypted \texttt{.text}};
\draw[arr] (tag.south)  -- ++(0,-6mm)
    node[lbl, below, align=left]
        {Final ELF: \texttt{.text} (encrypted) $+$ \texttt{.text.tag}};
 
\draw[arr] ($(link.west)+(-20mm,0)$) -- (link.west)
    node[lbl, midway, above, yshift=0.5mm, xshift=-6mm]
        {Libraries (\texttt{.lib}, \texttt{.a})};
\node[stp, right=1mm of pre]
    {\textbf{Step 1:} \textcolor{blue}{Preprocessor (\texttt{cpp})}};
\node[stp, right=1mm of comp]
    {\textbf{Step 2:} \textcolor{blue}{Compiler (\texttt{gcc}, \texttt{g++})}};
\node[stp, right=1mm of asm]
    {\textbf{Step 3:} \textcolor{blue}{Assembler (\texttt{as})}};
\node[stp, right=1mm of link]
    {\textbf{Step 4:} \textcolor{blue}{Linker (\texttt{ld})}};
\node[stp, right=1mm of enc]
    {\textbf{Step 5:} \textcolor{blue}{Encrypt \texttt{.text}}};
\node[stp, right=1mm of tag]
    {\shortstack[l]{\textbf{Step 6:} \textcolor{blue}{Write tags to}\\
                    \hphantom{\textbf{Step 6:} }\textcolor{blue}{\texttt{.text.tag}}}};
 
\end{tikzpicture}%
}
\caption{Modified RISC-V toolchain with instruction (\texttt{.text}) encryption and authentication tag generation (in \texttt{.text.tag}).}
\label{fig:toolchain}
\end{figure}

 
A design goal of \schemename{} is that programs are built with the standard flow, and encryption and tag generation are post-link transformations over the resulting binaries (\Cref{fig:toolchain}).  \schemename{} requires fixed-width 32-bit instructions (\Cref{subsec:single}). A GCC toolchain targeting \texttt{rv32im} was built with compressed (16-bit) instructions disabled. Applications and standard libraries were compiled with it; the absence of compressed instructions was verified by disassembly across the software stack. Two back-ends then apply the same per-word ASCON-128a encryption to the linked output.
 
\emph{ELF back-end.} A modified \texttt{objcopy} post-processes compiled ELF binaries, encrypting the \texttt{.text} section in place as independent 4-byte ciphertext blocks and leaving the binary's layout untouched. Authenticated mode additionally stores one 16-byte ASCON-128a tag per instruction in a dedicated \texttt{.text.tag} section while keeping \texttt{.text} at its original density; this requires no branch patching, preserves ISA compliance, permits parallel power-of-two-aligned fetches of instructions and tags, and confines all cryptographic logic to a single toolchain file with no BFD dependencies. A custom linker script defines two separate \texttt{PT\_LOAD} program headers---one for \texttt{.text} (read/execute) and one for \texttt{.text.tag} (read-only)---and reserves address space for the tag section as a \texttt{NOBITS} placeholder of size \texttt{SIZEOF(.text) $\times$ 4}, i.e., one 16-byte tag per 4-byte instruction, realising the $5\times$ instruction-memory cost established in \Cref{subsec:single}. The modified \texttt{objcopy} then performs two passes: it encrypts each instruction in place, collects the resulting tags into a separate buffer, writes the encrypted instructions back into \texttt{.text}, and writes the tag buffer into \texttt{.text.tag}, which the BFD library promotes from \texttt{NOBITS} to \texttt{PROGBITS}; section-flag corrections in \texttt{setup\_section()} ensure \texttt{.text.tag} carries correct allocate-only flags.
 
\emph{FPGA back-end.} For FPGA targets, an offline C++ tool applies the same per-word encryption directly to the NEORV32 instruction-memory image (\texttt{neorv32\_imem\_image.vhd}), emitting in a single pass the encrypted IMEM image and a matching TAGMEM package (encryption-only designs omit the latter). The back-ends differ only in scope: IMEM-level integration encrypts both instructions and read-only data (\texttt{.rodata}), whereas frontend integration encrypts only the instruction section (\texttt{.text}).
 
\subsection{Spike Functional Reference Model}\label{subsec:spike}
 
To provide a reference independent of the RTL, we extended the Spike RISC-V ISA simulator~\cite{riscvisasim} with ASCON-128a support from the reference C implementation~\cite{asconc}, first for encryption-only execution and then for authenticated execution. Spike's ELF loader records the encrypted region's virtual-address bounds, and the MMU's \texttt{refill\_icache()} invokes ASCON-128a decryption on a cache miss for fetches within it, before decode. Unencrypted binaries remain backward-compatible; decryption activates only when a valid encrypted \texttt{.text} region is detected at load time. In authenticated mode, Spike maps \texttt{.text.tag} into simulated memory as an ordinary \texttt{PT\_LOAD} segment, requiring no loader modification. The fetch unit locates the tag of any instruction using the address formula \texttt{tag\_addr = TAG\_BASE + ((PC - TEXT\_BASE) >> 2) $\times$ 16}, where \texttt{TAG\_BASE} is read dynamically from the \texttt{\_tag\_start} ELF symbol at load time, reducing tag addressing to a single shift and add. Upon a cache miss, the fetch unit retrieves the 4-byte encrypted instruction and its corresponding 16-byte tag, invokes ASCON-128a decryption with authentication, and raises a fetch fault/illegal opcode if tag verification fails---mirroring, the pre-decode fault semantics of the hardware designs (ISA level) (\Cref{subsec:single}).
 
\subsection{Experimental Setup and Verification}\label{subsec:setup}
 
Functional verification used the modified Spike RISC-V ISA simulator (version~1.1.1), with GNU Binutils 2.46, including the GNU assembler and \texttt{objcopy}, and GCC 15.2.0 (\texttt{riscv32-unknown-elf}). Our hardware prototype targets the Digilent Nexys A7-100T board (Xilinx Artix-7 XC7A100T), with on-chip block RAM (BRAM) for code and data and UART for runtime interaction, synthesised and implemented with Vivado ML Edition~2025.2. Two C++ suites were used across simulation and hardware: \texttt{hello\_world} (610 static instructions in \texttt{.text}; 822 executed) and Dhrystone with 10{,}000 iterations (1{,}214 static in \texttt{.text}; 4{,}541{,}606 executed). On the baseline processor these require 3{,}995 and 18{,}796{,}356 clock cycles, respectively, excluding input/output cycles.
 
 
The flow was verified end-to-end at four levels. At the software level, a companion C++ tool performed decryption of every 4-byte ciphertext with its 16-byte tag and matched the original plaintext image bit-exactly, with zero authentication failures. At the ISA level, the encrypted, authenticated binaries executed correctly under the extended Spike, exercising the loader, tag addressing, and authenticated-fetch path independently of the RTL. 
At the simulation level, GHDL (VHDL-2008) runs of encrypted, authenticated \texttt{hello\_world} and Dhrystone confirmed correct decryption and authentication across all 822 and 4{,}541{,}606 executed instruction words, respectively, reproducing the expected UART output without errors. In hardware, the designs were synthesised, implemented and executed on the target FPGA. These checks were complemented by negative testing: deliberate corruption of instruction words in IMEM and tags in TAGMEM caused every affected word to fail authentication, deasserting \texttt{tag\_ok} and correctly raising an illegal instruction exception in both GHDL simulation and on the FPGA.
 
\section{Evaluation}\label{sec:eval}

\subsection{Synthesis Results}

\Cref{tab:baseCPU} presents synthesis results for the baseline NEORV32 alongside the single-cycle ASCON-128a decryptors in tagless and authenticated configurations. The evaluated NEORV32 instance is a single-core design with the compressed-instruction (C) and code-size-reduction (Zcb) extensions disabled and 32\,KB/64\,KB instruction/data memories. The ASCON dominates hardware cost in both configurations, driving area, critical-path delay, and power. Notably, power consumption of the authenticated design is $\approx$4.4$\times$ the baseline, while LUT usage is $\approx$8.3$\times$ NEORV32.

\begin{figure}[t]
\centering
\begin{tikzpicture}
\begin{axis}[
  ybar,
  bar width=5pt,
  width=\linewidth,
  height=5.8cm,
  ymin=0, ymax=11,
  ytick={0,2,4,6,8,10},
  symbolic x coords={FEAUDC, MCP12, MCP6, MCP4, MCP3, MCP2, MCP1},
  xtick=data,
  xticklabel style={font=\footnotesize},
  yticklabel style={font=\footnotesize},
  ylabel={Overhead ($\times$ baseline)},
  ylabel style={font=\footnotesize},
  enlarge x limits=0.10,
  ymajorgrids,
  grid style={thin, gray!30},
  tick align=outside,
  axis line style={gray!60},
  legend style={
    at={(0.5,1.02)}, anchor=south,
    font=\footnotesize, legend columns=3,
    column sep=6pt, draw=none,
  },
  legend cell align=left,
legend image code/.code={
  \draw[#1] (0cm,-0.1cm) rectangle (0.3cm,0.1cm);
},
]

\addplot[ybar, draw=black, fill=black!80] coordinates {
  (FEAUDC, 10.0)
  (MCP12,  4.70)
  (MCP6,   2.356)
  (MCP4,   1.478)
  (MCP3,   1.156)
  (MCP2,   1.0)
  (MCP1,   1.0)
};
\addplot[ybar, draw=black, fill=black!25] coordinates {
  (FEAUDC, 1.0)
  (MCP12,  2.119)
  (MCP6,   2.602)
  (MCP4,   3.085)
  (MCP3,   3.568)
  (MCP2,   4.535)
  (MCP1,   7.434)
};
\addplot[ybar, draw=black, pattern=crosshatch dots, pattern color=black] coordinates {
  (FEAUDC, 9.348)
  (MCP12,  4.990)
  (MCP6,   3.315)
  (MCP4,   2.631)
  (MCP3,   2.217)
  (MCP2,   1.900)
  (MCP1,   1.559)
};
\legend{CCT, CPI, LUT}
\draw[dashed, black!60]
  ({rel axis cs:0,0} |- {axis cs:FEAUDC,1})
  -- ({rel axis cs:1,0} |- {axis cs:FEAUDC,1});
\end{axis}
\end{tikzpicture}

{\small (a) Micro-architectural factors.\vspace{2mm}}
\begin{tikzpicture}
\begin{axis}[
  ybar,
  bar width=5pt,
  width=\linewidth,
  height=5.8cm,
  ymode=log,
  log origin=infty,
  ymin=1, ymax=130,
  ytick={1,2,5,10,20,50,100},
  yticklabels={1,2,5,10,20,50,100},
  symbolic x coords={FEAUDC, MCP12, MCP6, MCP4, MCP3, MCP2, MCP1},
  xtick=data,
  xticklabel style={font=\footnotesize},
  yticklabel style={font=\footnotesize},
  ylabel={Overhead ($\times$ baseline)},
  ylabel style={font=\footnotesize},
  enlarge x limits=0.10,
  ymajorgrids,
  grid style={thin, gray!30},
  tick align=outside,
  axis line style={gray!60},
  legend style={
    at={(0.5,1.02)}, anchor=south,
    font=\footnotesize, legend columns=3,
    column sep=6pt, draw=none,
  },
  legend cell align=left,
  legend image code/.code={
    \draw[#1] (0cm,-0.1cm) rectangle (0.3cm,0.1cm);
  },  
]
\addplot[ybar, draw=black, pattern=north east lines, pattern color=black!75] coordinates {
  (FEAUDC, 10.0)
  (MCP12,  9.958)
  (MCP6,   6.129)
  (MCP4,   4.559)
  (MCP3,   4.124)
  (MCP2,   4.585)
  (MCP1,   7.434)
};
\addplot[ybar, draw=black, fill=black!2] coordinates {
  (FEAUDC, 8.0)
  (MCP12,  3.230)
  (MCP6,   3.425)
  (MCP4,   4.617)
  (MCP3,   4.167)
  (MCP2,   2.258)
  (MCP1,   1.492)
};
\addplot[ybar, draw=black, pattern=crosshatch, pattern color=black] coordinates {
  (FEAUDC, 79.7)
  (MCP12,  32.16)
  (MCP6,   20.99)
  (MCP4,   21.05)
  (MCP3,   17.18)
  (MCP2,   10.36)
  (MCP1,   11.09)
};
\legend{Performance, Power, EPI}
\end{axis}
\end{tikzpicture}

{\small (b) System-level factors.}
\caption{Normalised metrics relative to the NEORV32 baseline using Dhrystone. The dashed line in~(a) marks $1\times$.}
\label{fig:barchart9}
\vspace{-2mm}
\end{figure}
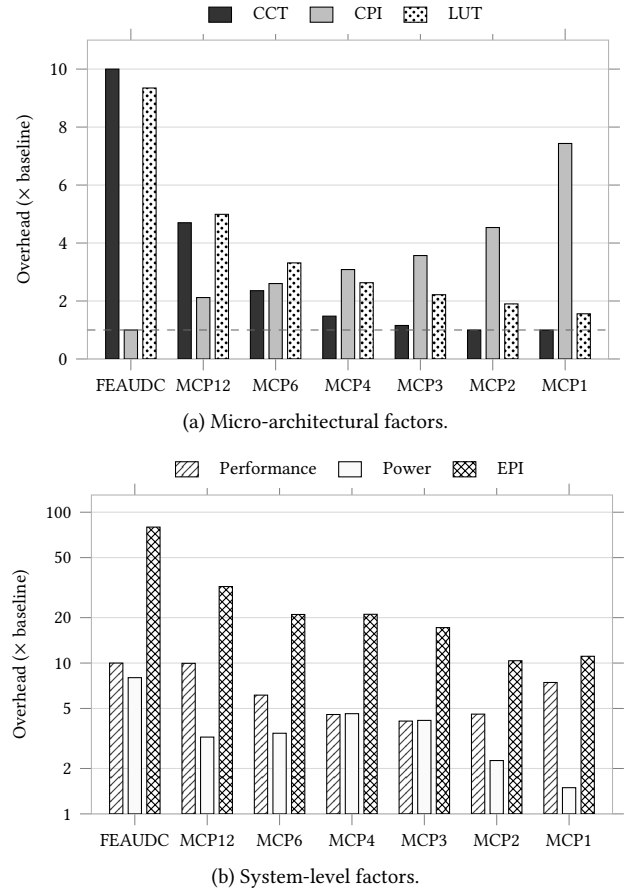

\subsection{Single-Cycle Design Results}

\Cref{fig:FullyCombASCON} compares the four single-cycle secure configurations against the baseline. The four configurations are: (1) tagless ASCON-128a decryption integrated into the instruction memory wrapper (\texttt{IMDEC}); (2) fully authenticated ASCON-128a decryption integrated into the instruction memory wrapper (\texttt{IMAUDC}); (3) tagless ASCON-128a decryption integrated into the CPU frontend (\texttt{FEDEC}); and (4) fully authenticated ASCON-128a decryption integrated into the CPU frontend (\texttt{FEAUDC}). The capability rows expose a security-relevant distinction: IMEM-level integration additionally protects read-only data but places plaintext on the processor bus, whereas frontend integration keeps ciphertext on the bus, denying the instruction stream to an attacker with bus-monitoring capability. CPI is invariant across all configurations, since the fully combinational (authenticated) decryption completes within a single, elongated clock cycle and introduces no stalls or wait states; each configuration's performance overhead therefore equals its clock-cycle-time ratio to the baseline. Adding authentication roughly doubles area and delay over tagless decryption, as two ASCON 12-round permutation instances are required, and \texttt{IMAUDC}/\texttt{FEAUDC} each consume 8 additional BRAM tiles for the TAGMEM holding one 128-bit tag per instruction word. The baseline was synthesised and implemented under a 100\,MHz timing constraint; the four single-cycle secure configurations under 10\,MHz, owing to the ASCON critical path.

\begin{table}
\caption{Synthesis results.}
\label{tab:baseCPU}
\begin{center}
\footnotesize
\begin{tabular}{r|ccccc}
\toprule
\textbf{Design} & \textbf{LUTs} & \textbf{FFs} & \textbf{\shortstack{Power \\ (mW)}} & \textbf{\shortstack{Delay \\ (ns)}} & \textbf{\shortstack{Block\\RAM Tile}} \\
\midrule
NEORV32 (Base) & 1867 & 1428 & 120 & 9 & 19 \\
ASCON-128a (Enc.-only) & 7590 & 0 & 336 & 29.8 & 0 \\
ASCON-128a (Auth.) & 15621 & 0 & 533 & 88 & 0 \\
\bottomrule
\end{tabular}
\end{center}
\end{table}
\begin{table}
\caption{Single-cycle design results (performance overhead given as $N \times$ vs.~NEORV32 baseline).}
\begin{center}
\footnotesize
\begin{tabular}{lcccc}
\toprule
\textbf{Metrics} & \textbf{IMDEC} & \textbf{IMAUDC} & \textbf{FEDEC} & \textbf{FEAUDC} \\
\midrule
Authentication & \xmark & \cmark & \xmark & \cmark \\
Clock Cycle Time & $5.4\times$ & $10.2\times$ & $4.9\times$ & $10\times$ \\
CPI & $1\times$ & $1\times$ & $1\times$ & $1\times$ \\
LUTs Overhead & $4.8\times$ & $9.3\times$ & $4.8\times$ & $9.3\times$ \\
Performance Overhead & $5.4\times$ & $10.2\times$ & $4.9\times$ & $10\times$ \\
Power Overhead & $4.3\times$ & $7.9\times$ & $4.3\times$ & $8\times$ \\
\bottomrule
\end{tabular}
\label{fig:FullyCombASCON}
\end{center}
\end{table}

\subsection{Multi-Cycle Design Results}
\label{subsec:multicycle-results}

\Cref{fig:barchart9} characterises the multi-cycle design space using
Dhrystone for \texttt{FEAUDC}---the single-cycle upper bound
of the design space---and the six multi-cycle configurations,  normalised to the NEORV32 baseline.
\Cref{fig:barchart9}(a) reports the three microarchitectural factors:
clock cycle time (CCT), CPI, and LUT count.
\Cref{fig:barchart9}(b) reports the three system-level overheads: performance, power, and energy per instruction (EPI).
Because the executed instruction count is invariant across
configurations while both the clock period and the CPI vary, the
performance overhead of each configuration is the product of its
normalised CCT and CPI; EPI is in turn the product of the normalised
power and performance overheads.
EPI is invariant to the per-design clock frequency, enabling the comparison of configurations synthesised under different timing constraints.

Each design was implemented at the highest clock frequency at which it
closed timing: the baseline, \texttt{MCP1}, and \texttt{MCP2} at
100\,MHz; \texttt{MCP3} at 90\,MHz; \texttt{MCP4} at 70\,MHz;
\texttt{MCP6} at 25\,MHz; and \texttt{MCP12} and \texttt{FEAUDC} at
10\,MHz, reflecting their progressively longer critical paths.
Power estimates are activity-aware: switching activity captured during
post-implementation functional simulation was exported in the Switching
Activity Interchange Format (SAIF) and annotated into the power-analysis
flow, and each design's clock period derives from its
post-implementation critical-path delay.
Since the configurations are therefore synthesised under different
timing constraints, raw power is not directly comparable across
designs; the EPI metric normalises for this.

\Cref{fig:barchart9}(a) exposes the central trade-off of the design
space.
Folding the ASCON-p[12] permutation across more clock cycles places
fewer rounds, and hence less combinational logic, in each stage, so
the critical path shortens monotonically (CCT falls from $10\times$ at
\texttt{FEAUDC} to $1\times$ at \texttt{MCP2} and \texttt{MCP1}) and the
decryptor shrinks (from $9.3\times$ to $1.6\times$ LUTs), while the CPI
rises monotonically (from $1\times$ to $7.4\times$) as authenticated
decryption completes over a greater number of cycles.
The overheads in \Cref{fig:barchart9}(b) show that no single
configuration dominates: the design space contains three distinct
optima.
\texttt{MCP1} minimises area ($1.6\times$ LUTs), power ($1.5\times$),
and CCT ($1\times$); \texttt{MCP3} minimises performance overhead
($4.1\times$); and \texttt{MCP2} minimises EPI overhead ($10.4\times$).
Moreover, folding the permutation lowers EPI by roughly an order of magnitude relative to the single-cycle \texttt{FEAUDC} ($79.7\times$), demonstrating that the multi-cycle designs are smaller and fundamentally more energy-efficient per executed instruction.

\subsection{Security Analysis}

\schemename{} draws the secrecy of stored code directly from ASCON-128a under a CPU-bound key that is never exposed. Nonce uniqueness is required for ASCON-128: within a single code image, the base-nonce-plus-address construction ensures a distinct nonce per instruction word. Across re-encryptions (e.g., firmware updates) each image must be produced under a fresh key or fresh base nonce, which we assume the provisioning process enforces. Identical instructions, and instructions sharing fields such as opcodes, thus yield unrelated ciphertexts. An adversary confined to the encrypted memory is unable to access plaintext code. 

Nevertheless, an adversary who cannot read the code can still modify it, and a corrupted word may itself decode into an executable instruction. Moreover, encrypted IMEM does not hide which addresses are fetched, so access and timing patterns may still leak coarse program structure. The latter is outside our scope; the former is closed by binding each ciphertext word to its stored address through the address-derived nonce and to a tag checked as the word is decrypted. Any change to the ciphertext, its tag, or its location makes the check fail, and an exception is raised deterministically  before it is executed. This is a sharper guarantee than the implicit schemes~\cite{werner2018sponge,gousselot2025code}, where a tampered word decrypts to a pseudo-random value caught only when it decodes illegally, by which point several instructions may already have executed. Bit-flips in IMEM or TAGMEM fail authentication and halt execution; injected code carries no valid tag and is rejected before decode, while relocation and splicing fail because the nonce is tied to the fetch address. \schemename{} also defeats the Cipher Instruction Search Attack~\cite{kuhn1998cipher}: since no ciphertext can verify without the key, an adversary cannot iteratively probe the decryptor to assemble a working instruction. 

\section{Conclusion and Future Work}

We introduced a design exploration of \schemename{}, which performs instruction-level authenticated decryption on RISC-V, protecting stored code against injection, tampering, and static reverse engineering threats. We presented and integrated seven decryptor implementations using ASCON-128a---single-cycle down to one round per cycle---at the NEORV32 IMEM wrapper and CPU frontend in tagless and authenticated configurations, which were validated in an extended Spike ISA simulator, in GHDL simulation, and on an Artix-7 FPGA. Relative to the baseline, LUT, performance, power, and EPI overheads span $9.3\times$--$1.6\times$, $10\times$--$4.1\times$, $8\times$--$1.5\times$, and $80\times$--$10.4\times$, with authenticated execution requiring $5\times$ instruction memory for per-instruction 128-bit tags. Explicit, pre-execution authenticated execution is thus feasible on lightweight embedded RISC-V cores, and this work constitutes, to our knowledge, the first FPGA-validated power--performance--area--energy design-space exploration of instruction-level ASCON-128a authenticated decryption. Future areas of research include supporting CFI enforcement, authenticated encryption of data memory, and compressed-instruction support.

\ifanon\else
\begin{acks}
This work was supported by the \grantsponsor{EPSRC}{UK Engineering and Physical Sciences Research Council}{https://www.ukri.org/councils/epsrc/} via the `Chameleon' project (\grantnum{EPSRC}{EP/Y030168/1}).
\end{acks}
\fi

\bibliographystyle{ACM-Reference-Format}
\bibliography{refs}

\end{document}